\documentclass[twocolumn,showpacs,amsmath,amssymb]{revtex4}

\usepackage{graphicx}

\begin{document}
\title{A new method for determining dipole-dipole energy in 1D and 2D systems}
\author{Ivan I. Naumov and Huaxiang Fu}
\affiliation{Department of Physics,  University of Arkansas,
Fayettville, AR 72701}
\date{\today}

\begin{abstract}
An alternative method for computing dipole-dipole interaction
energy in systems of 1D and 2D periodicity like nanowires,
nanotubes and thin films is presented. The approach is based on
the use of periodic Green's functions that satisfy Laplace's
equation and are analytically determined. The method, when
combined with short-ranged interaction as in effective
Hamiltonian, is suitable for studying finite-temperature
properties of low-dimensional ferroelectric systems.
\end{abstract}


\maketitle

\section{Introduction}

In the past decade it was shown that the finite-temperature
behaviour of ferroelectric systems like BaTiO$_3$ can be
successfully simulated by using the statistical mechanics of an
Effective Hamiltonian, which, in its turn, is based on  the
first-principles calculations of total energies for small
distortions of the high-temperature cubic structure. Such an
approach predicts sequence of phase transformations,
electromechanical responses and other finite-temperature
properties in a good agreement with experiment.

The conventional effective first-principles  Hamiltonian approach
is developed for the bulk systems infinitely repeated in all three
Cartesian directions (3D case) \cite{zhong}. Formally it can also
be applied to the systems which are effectively infinite only  in
one or two dimensions simply repeating them many times in finite
direction(s) (in this case sufficiently thick vacuum gap(s) must
be created between periodic replicas). Such a procedure, however,
inevitably leads to errors in   depolarizing electric fields and
in corresponding shape-dependent electrostatic energy
\cite{bengtsson,yeh,brodka, brodka2}. Consequently, in passing
from the systems with three- to that with two- and one-dimensional
periodicity the part of the effective Hamiltonian connected with
long-range dipole-dipole interactions should be modified.

The existing methods for rapid evaluation of dipole-dipole
interactions in 1D \cite{porto,brodka3} and 2D \cite{heyes1,
heyes2,jensen,grz} systems are based on Ewald type summation
technique exploiting the integral representation of the Gamma
function (or Euler's integral) and Poisson summation formula. This
technique leads to fast convergent  sums  in real and reciprocal
spaces, which, however, are rather bulky.

Here, we present a new elegant method for the treatment of
dipole-dipole interaction in partially periodic systems  which
leads to much simpler final expressions than the Ewald technique.
The method is based on using the $analytically$ determined
$periodic$ Green's function of the Laplace's equation
$\mathcal{G}(\mathbf{r'},\mathbf{r})$. The simplicity  of our
final expressions is partly  due to the fact  that we do not use
Ewald type transformations in 1D case at all, whereas in 2D case
we use it only for the special case when the the interacting
dipoles lay in the same plane. The other reason comes from the
fact that we perform the summation of dipole-dipole  interactions
(from dipoles located in different unit cells) entirely in
reciprocal space--   with the only exception for the 1D case when
the dipoles lay  along the line parallel to the periodicity
direction.

\section{Dipole-dipole interactions in systems with 1D and 2D periodicity}

We regard the systems with 1D and 2D periodicity as infinite  in
one  and two of the three Cartesian directions correspondingly.
They can be obtained by replicating a "unit cell" an infinite
number of times along that directions. The dipole-dipole
interaction energy can be written as
\begin{eqnarray}
\mathcal{E}_{dip}=\frac{1}{2}\sum_{i\neq j}\bigg\{
\frac{\mathbf{d}(\mathbf{R}_{i})\cdot\mathbf{d}(\mathbf{R}_{j})}{R^{3}_{ij}}-
\frac{3\,[\mathbf{d}(\mathbf{R}_{i})\cdot\mathbf{R}_{ij}]
[\mathbf{d}(\mathbf{R}_{j})\cdot\mathbf{R}_{ij}
]}{R^{5}_{ij}}\bigg\}\nonumber\\
=\frac{1}{2}\sum_{\alpha \beta}\sum_{i\neq j}D_{\alpha
\beta}(\mathbf{R}_{i}-\mathbf{R}_{j})
d_{\alpha}(\mathbf{R}_{i})d_{\beta}(\mathbf{R}_{j}),\qquad
\label{eq:1}
\end{eqnarray}
where
$\mathbf{R}_{ij}=\mathbf{R}_{i}-\mathbf{R}_{j},\:R_{ij}=|\mathbf{R}_{ij}|$
is the distance between dipoles, $d_{\alpha}(\mathbf{R}_{i})$ is
the $\alpha$-component ($\alpha=x,y,z$) of the dipole moment at
the site $i$  and
\begin{equation}
D_{\alpha\beta}(\mathbf{R}_{i}-\mathbf{R}_{j})=-
\lim_{\mathbf{r}\,\to\,0} \frac{\partial}{\partial
r_{\alpha}}\frac{\partial}{\partial r_{\beta}}\bigg(\frac{1}
{\mid \mathbf{r}-\mathbf{R}_{i}+\mathbf{R}_{j}\mid}\bigg).
\label{eq:2}
\end{equation}
Due to periodicity each vector $\mathbf{R}_{i}$ is represented as
$\mathbf{r}_{i}-\mathbf{R}_{\parallel}$ where $\mathbf{r}_{i}$ is
the dipole position inside the 0-th unit cell and
$\mathbf{R}_{\parallel}$ is an appropriate vector from the
infinite number of vectors forming 1D or 2D lattice. Accordingly,
we may rewrite Eq.(\ref{eq:1}) as
\begin{equation}
\mathcal{E}_{dip}=\frac{N}{2}\sum_{\alpha\beta,\mathbf{R}_{\parallel}}
\sum^{\qquad\prime}_{ij}D_{\alpha
\beta}(\mathbf{r}_{i}-\mathbf{r}_{j}+
\mathbf{R}_{\parallel})d_{\alpha}(\mathbf{r}_{i})d_{\beta}(\mathbf{r}_{j}),
 \label{eq:3}
\end{equation}
where the summation over $i,j$ runs only  inside  the 0-th unit
cell, the prime  means that the term with
$\mathbf{r}_{i}=\mathbf{r}_{j}$ in the case of
$\mathbf{R}_{\parallel}=0$ must be omitted, $N$ is the number of
unit cells allowed  to tend to infinity. We are interested in
$\mathcal{E}$ per unit cell, so $N$ in Eq.(\ref{eq:3}) can be
omitted. Using this we can recast Eq.(\ref{eq:3}) as
\begin{equation}
\mathcal{E}_{dip}= -\lim_{\mathbf{r}\,\to\,0}\frac{1}{2}\sum_{
\alpha\beta, ij }^{\qquad\prime} \frac{\partial}{\partial
r_{\alpha}}\frac{\partial}{\partial
r_{\beta}}\mathcal{G}(\mathbf{r,r'})\mid_
{\mathbf{r'}=\mathbf{r}_{i}-\mathbf{r}_{j}}
d_{\alpha}(\mathbf{r}_{i})d_{\beta}(\mathbf{r}_{j}),
 \label{eq:4}
\end{equation}
where
\begin{equation}
\mathcal{G}(\mathbf{r,r'})=\sum_{\mathbf{R}_{\parallel}}\bigg(\frac{1}
{\mid \mathbf{r}-\mathbf{r'}+\mathbf{R}_{\parallel}\mid}\bigg)
\end{equation}
 \label{eq:5}
is nothing but the  periodic Green's function of Laplace's
equation satisfying the point source equation
\begin{equation}
\nabla^2_{r}\mathcal{G}(\mathbf{r,r'}) =
-4\pi\sum_{\mathbf{R}_{\parallel}}\delta(\mathbf{r-r'+\mathbf{R}_{\parallel}}),
\label{eq:6}
\end{equation}
and the translation symmetry
\begin{equation}
\mathcal{G}(\mathbf{r,r'})=\mathcal {G}(\mathbf{r}+\mathbf{R}_{
\parallel}, \mathbf{r'}+\mathbf{R'}_{\parallel}).
 \label{eq:7}
\end{equation}
Now our task  is to evaluate $\mathcal{G}(\mathbf{r,r'})$  and
then $\mathcal{E}_{dip} $ for 1D and 2D cases according to
(\ref{eq:4}).

\subsection{1D case}
Let $z$-axis to be along the infinite dimension, then  all the
vectors $\mathbf{R}_{\parallel}$ lay in it. We represent each
vector $\mathbf{r}$ as decomposed into two components
$\{\boldsymbol{\rho},z\}$, where $\boldsymbol{\rho}$ is the
projection of the $\mathbf{r}$ on the $(x,y)$ plane. Accordingly,
the Dirac functions $
\delta(\mathbf{r-r'+\mathbf{R}_{\parallel}})$ in the right-hand
side of Eq.(\ref{eq:6}) become
$\delta(\boldsymbol{\rho}-\boldsymbol{\rho}')\,
\delta(\mathbf{z-z'+R_{\parallel}})$. The 2D and the sum of 1D
Dirac functions can be expressed as  Fourier integral and sum
correspondingly:
\begin{equation}
\delta(\boldsymbol{\rho}-\boldsymbol{\rho}')=
\frac{1}{(2\pi)^{2}}\int e^{i
\mathbf{k}_{\perp}\cdot(\boldsymbol{\rho}-\boldsymbol{\rho}')}d\mathbf{k}_{\perp},
 \label{eq:8}
\end{equation}
\begin{equation}
\sum_{\mathbf{R}_{\parallel}}\delta(\mathbf{z-z'+R_{\parallel}})=
\frac{1}{a}\sum_{\mathbf{G}_{\parallel}}e^{i\mathbf{G}_{\parallel}\cdot(\mathbf{z-z'})},
\label{eq:9}
\end{equation}
where $\mathbf{k}_{\perp}$ is the 2D wave-vector perpendicular to
the $z$-direction, $a$ is the period along this direction, and
$\mathbf{G}_{\parallel}$ are the reciprocal lattice vectors
corresponding to the 1D lattice of repeated cells. We shall look
for the Green's function in the form
\begin{equation}
\mathcal{G}(\mathbf{r,r'})=\sum_{\mathbf{G}_{\parallel}}\int
g(\mathbf{k}_{\perp},\mathbf{G}_{\parallel})
e^{i(\mathbf{k}_{\perp}+\mathbf{G}_{\parallel})\cdot
(\mathbf{r}-\mathbf{r}')}d\mathbf{k}_{\perp}
\label{eq:10}
\end{equation}
Inserting (\ref{eq:8}), (\ref{eq:9}), and (\ref{eq:10}) into
Eq.(\ref{eq:6}) we obtain\cite{morse,jackson}
\begin{eqnarray}
\mathcal{G}(\mathbf{r,r'})=\frac{1}{\pi
a}\sum_{\mathbf{G}_{\parallel}} \int \frac{e^{i
\mathbf{k}_{\perp}\cdot(\boldsymbol{\rho}-\boldsymbol{\rho}')}
e^{i\mathbf{G}_{\parallel}\cdot(\mathbf{z-z'})}}
{\mathbf{G}^{2}_{\parallel}+\mathbf{k}^{2}_{\perp}}d\mathbf{k}_{\perp}\nonumber\\
=\frac{2}{a}\sum_{\mathbf{G}_{\parallel}}K_{0}\big(G_{\parallel}
|\boldsymbol{\rho}-\boldsymbol{\rho}'|\big)\,
e^{i\mathbf{G}_{\parallel}\cdot(\mathbf{z-z'})},
 \label{eq:11}
\end{eqnarray}
where $K_{0}(x)$ is the  0-th order modified Bessel's function of
the imaginary argument. Since the Bessel function $K_{0}(x)$
decays exponentially for large argument $x$, the series over
$\mathbf{G}_{\parallel}$ (\ref{eq:11}) converge much faster than
the direct lattice sum (\ref{eq:1}). This is valid only for the
contributions to $\mathcal{E}_{dip}$ with
$\boldsymbol{\rho}_{i}\neq \boldsymbol{\rho}_{j}$. As to
contributions coming from the dipole-dipole interactions with
$\boldsymbol{\rho}_{i}= \boldsymbol{\rho}_{j}$ or, in other
words,  from the chains of dipoles parallel to the z-axis, they
can be easily calculated in the real space; the corresponding 1D
sums are rapidly convergent like $1/n^{3}$. Moreover, doing so
one can automatically solve the problem of excluding of the
"self-interaction" term from the sum (\ref{eq:3}).

Substituting (\ref{eq:11}) into Eq.(\ref{eq:4}) and taking into
account that $K'_{0}(x)=-K_{1}(x)$ and
$K''_{0}(x)=K_{2}(x)-K_{1}(x)/x$ one can easily obtain
\begin{eqnarray}
&&\mathcal{E}_{dip}=\frac{1}{a}\sum_{\mathbf{G}_{\parallel}}
\sum_{ij}^{\qquad\prime}
G^{2}_{\parallel}\,\cos(\mathbf{G}_{\parallel}\cdot
\mathbf{z}_{ij})\nonumber\\
&&\times\bigg\{K_{0}\big(G_{\parallel}\rho_{ij}\big) \:
d_{z}(\mathbf{r}_{i})d_{z}(\mathbf{r}_{j})\nonumber\\
&&+\frac{1}{G_{\parallel}\,\rho_{ij}}
K_{1}\big(G_{\parallel}\rho_{ij}\big)\big[d_{x}(\mathbf{r}_{i})
d_{x}(\mathbf{r}_{j})+d_{y}(\mathbf{r}_{i})
d_{y}(\mathbf{r}_{j})\big]\nonumber\\
&&-\frac{1}{\rho_{ij}^2}K_{2}\big(G_{\parallel} \rho_{ij}\big)\,
\big[\boldsymbol{\rho}_{ij}\cdot\mathbf{d}
(\mathbf{r}_{i})\big]\,\big[\boldsymbol{\rho}_{ij}\cdot\mathbf{d}
(\mathbf{r}_{j})\big]\bigg\}\nonumber\\
&&-\frac{1}{a}\sum_{\mathbf{G}_{\parallel}}\sum_{ij}^{\qquad\prime}\:
G_{\parallel}\,\sin(\mathbf{G}_{\parallel}\cdot \mathbf{z}_{ij})
\:K_{1}\big(G_{\parallel} \rho_{ij}\big)\,\rho_{ij}^{-1}\nonumber\\
&&\times\bigg\{\big[\mathbf{G}_{\parallel}\cdot\mathbf{d}(\mathbf{r}_{i})\big]
\big[\boldsymbol{\rho}_{ij}\cdot\mathbf{d} (\mathbf{r}_{j})\big]+
\big[\mathbf{G}_{\parallel}\cdot\mathbf{d}(\mathbf{r}_{j})\big]
\big[\boldsymbol{\rho}_{ij}\cdot\mathbf{d} (\mathbf{r}_{i})\big]\bigg\}\nonumber\\
&&+\frac{1}{2a^{3}}\sum_{ij}\big[d_{x}(\mathbf{r}_{i})\,d_{x}(\mathbf{r}_{j})
+d_{y}(\mathbf{r}_{i})\,d_{y}(\mathbf{r}_{j})-
2d_{z}(\mathbf{r}_{i})\,d_{z}(\mathbf{r}_{j})\big]\nonumber\\
&&\times f(|z_{ij}|).
 \label{eq:12}
\end{eqnarray}
Here, $G_{\parallel}=|\mathbf{G}_{\parallel}|,\,\mathbf{z}_{ij}=
\mathbf{z}_{i}-\mathbf{z}_{j},\,
\boldsymbol{\rho}_{ij}=\boldsymbol{\rho}_{i}-
\boldsymbol{\rho}_{j},\, \rho_{ij}=|\boldsymbol{\rho}_{ij}|$, and
\begin{equation}
f(|z_{ij}|)=\sum_{n=-\infty}^{\infty\quad\prime}
\bigg|n+\frac{z_{ij}}{a}\bigg|^{-3}, \label{eq:13}
\end{equation}
where the prime at the sum means  that  the term  $n=0$ is to  be
excluded for the case  $i=j$. The last sum in (\ref{eq:12})
describes the  contribution to the energy associated with the
chains parallel to the $z$-axis; this contribution is separated
from the first two sums marked by the primes. Note that the
function $f(|z_{ij}|)$ is periodical in real space with period
$a$. It is also worth noticing that the term $G_{\parallel}=0$
does contribute to the sum(\ref{eq:12}). Taking into consideration
that as  $x \to 0$
 $K_{0}(x)\to -ln(x)$, $K_{1}(x)\to 1/x$, and $K_{2}(x)\to
2/x^{2}$, we find
\begin{eqnarray}
\mathcal{E}_{dip}(G_{\parallel}=0)=&&\frac{1}{a}\sum_{i
j}^{\qquad\prime}\rho_{ij}^{-2}\bigg\{ d_{x}(\mathbf{r}_{i})
d_{x}(\mathbf{r}_{j})+d_{y}(\mathbf{r}_{i})
d_{y}(\mathbf{r}_{j})\nonumber\\
&&-2\rho_{ij}^{-2}\big[\boldsymbol{\rho}_{ij}\cdot\mathbf{d}
(\mathbf{r}_{i})\big]\,\big[\boldsymbol{\rho}_{ij}\cdot\mathbf{d}
(\mathbf{r}_{j})\big]\bigg\}. \label{eq:14}
\end{eqnarray}
If all $\mathbf{d} (\mathbf{r}_{i})\parallel$ to $z$, this
contribution turns to zero. In the macroscopic limit it describes
    the energy connected with the depolarizing field.
\subsection{2D case}
Consider a slab or thin film with normal along the z-direction,
each layer of which representing  an infinite  array of electric
dipoles in the  $(x,y)$ plane. For this geometry the vectors
$\mathbf{R}_{\parallel}$ form a 2D lattice parallel to this
plane. The corresponding Green's function, as easily to show
acting similar to the previous case and using the
methods\cite{jackson, arfken}, is
\begin{eqnarray}
\mathcal{G}(\mathbf{r,r'})=\frac{2}{S}\sum_{\mathbf{G}_{\parallel}}
\int\frac{e^{i\mathbf{k}_{z}\cdot(\mathbf{z-z}')}
e^{i\mathbf{G}_{\parallel}\cdot(\boldsymbol{\rho}-\boldsymbol{\rho}')}}
{\mathbf{G}^{2}_{\parallel}+\mathbf{k}^{2}_{z}}d \mathbf{k}_{z}\nonumber\\
=\frac{2\pi}{S}\sum_{\mathbf{G}_{\parallel}}\frac{e^{-G_{\parallel}|z-z'|}
e^{i\mathbf{G}_{\parallel}\cdot(\boldsymbol{\rho}-\boldsymbol{\rho}')}}
{G_{\parallel}},
 \label{eq:15}
\end{eqnarray}
where  $S$ is the primitive unit sell area \footnote{In a
particular case when the 2D lattice is rectangular with the $a$
and $b$ periods along the $x$ and $y$ axis, $S=a\,b$.}, parallel
to the slab plane, $\mathbf{G}_{\parallel}$ are the reciprocal
lattice vectors corresponding to the 2D lattice and
$\boldsymbol{\rho}=\{x,y\}$. By inserting (\ref{eq:15}) into
(\ref{eq:4}) it can be shown that only the contributions to
$\mathcal{E}_{dip}$ with $z_{i}\neq z_{j}$ exponentially converge
as $G_{\parallel}$ increases (compare with the 1D case). The
terms with $z_{i}=z_{j}$ describing the dipole-dipole
interactions in the $z$- layers need additional care; we separate
these terms and denote them  as $A$. Now instead of
Eq.(\ref{eq:12}) we have
\begin{eqnarray}
&&\mathcal{E}_{dip}=A+\frac{\pi}{S}\sum_{\mathbf{G}_{\parallel}}
\sum_{i,j}^{\qquad\prime}G_{\parallel}\,
\exp(-G_{\parallel}\,|z_{ij}|)\nonumber\\
&&\times\bigg[B\,\cos(\mathbf{G}_{\parallel}\cdot\boldsymbol{\rho}_{ij})
+C\,\sin(\mathbf{G}_{\parallel}\cdot\boldsymbol{\rho}_{ij})
\frac{z_{ij}}{|z_{ij}|}\bigg], \label{eq:16}
\end{eqnarray}
where
\begin{equation}
B= -d_{z}(\mathbf{r}_{i})d_{z}(\mathbf{r}_{j})
+\frac{1}{G_{\parallel}^2}
\big[\mathbf{G}_{\parallel}\cdot\mathbf{d}
(\mathbf{r}_{i})\big]\big[\mathbf{G}_{\parallel}\cdot\mathbf{d}
(\mathbf{r}_{j})\big], \label{eq:17}
\end{equation}
\begin{equation}
C=-\frac{1}{G_{\parallel}}\bigg\{\big[\mathbf{G}_{\parallel}\cdot\mathbf{d}
(\mathbf{r}_{i})\big]d_{z}(\mathbf{r}_{j})+
\big[\mathbf{G}_{\parallel}\cdot\mathbf{d}
(\mathbf{r}_{j})\big]d_{z}(\mathbf{r}_{i})\bigg\}.
 \label{eq:18}
\end{equation}
The prime at the second sum in (\ref{eq:16}) indicates  the terms
with $z_{ij}=0$ to be dropped. To evaluate these terms
 (collected  together in $A$) we apply the Ewald type
transformation to the Green's function (\ref{eq:15}) representing
it in both coordinate and reciprocal spaces:
\begin{eqnarray}
&&\mathcal{G}(\mathbf{r,r'})=
\sum_{\mathbf{R}_{\parallel}}\frac{\text{erfc}(\eta
|\mathbf{r-r'}+\mathbf{R}_{\parallel}|)}
{|\mathbf{r-r'}+\mathbf{R}_{\parallel}|}+\frac{2\sqrt{\pi}}{S}\nonumber\\
&&\times\sum_{\mathbf{G}_{\parallel}}
e^{i\mathbf{G}_{\parallel}\cdot(\boldsymbol{\rho}-\boldsymbol{\rho}')}
\int_{0}^{\eta}e^{-|z-z'|^{2}t^{2}}e^{-G_{\parallel}^{2}/4t^{2}
}t^{-2}dt.
 \label {eq:19}
\end{eqnarray}
Here $\eta$ is the Ewald parameter, and erfc is  the
complementary error function
\begin{equation}
\text{erfc}(x)={\frac{2}{\sqrt\pi}}\int_{x}^{\infty}\text{exp}(-t^{2})\,dt.
\label{eq:20}
\end{equation}
Using (\ref{eq:19}) and excluding the self-interaction term
corresponding to $\mathbf{r}_{i}=\mathbf{r}_{j}$ when
$\mathbf{R}_{\parallel}=0$, one obtains
\begin{eqnarray}
A =&& \frac{\pi}{S}\sum_{\mathbf{G}_{\parallel}}\sum_{ij}
G_{\parallel}\cos(\mathbf{G}_{\parallel}\cdot
\boldsymbol{\rho}_{ij})\nonumber\\
&&\times\Bigg\{\,\frac{1}{\sqrt{4\pi}}\:\Gamma\left(
 -\frac{1}{2},\,\frac{G_{\parallel}^{2}}{4\eta^{2}}
 \right) d_{z}(\mathbf{r}_{i})d_{z}(\mathbf{r}_{j})\nonumber\\
 &&+\frac{1}{G_{\parallel}^2}\:\text{erfc}\left(
 \frac{G_{\parallel}}{2\eta}
 \right)\big[\mathbf{G}_{\parallel}\cdot
 \mathbf{d}(\mathbf{r}_{i})\big]\big[\mathbf{G}_{\parallel}\cdot
 \mathbf{d} (\mathbf{r}_{j})\big]\Bigg\}
 \nonumber\\
&&-\sum_{i}\frac{2\eta^{3}\,|\mathbf{d}(\mathbf{r}_{i})|^{2}}{3\sqrt{\pi}},
 \label{eq:21}
\end{eqnarray}
where the summation over $i,j$ is constrained by the condition
$z_{ij}=0$, $\Gamma$ is the incomplete Gamma function\cite{arfken}
\begin{equation}
\Gamma(\alpha,x)=\int_{x}^{\infty}e^{-t}t^{\alpha-1}\,dt,
\label{eq:22}
\end{equation}
and the  Ewald parameter $\eta$ is presumed  to be large enough,
so that the real space summation can be entirely neglected. It is
interesting that the first sum in (\ref{eq:21}) contains the term
$G_{\parallel}=0$. Indeed, since
\begin{equation}
\lim_{G_{\parallel}\,\to\,0}\:\Gamma\left(
 -\frac{1}{2},\,\frac{G_{\parallel}^{2}}{4\eta^{2}}
 \right)=4\eta/G_{\parallel},
 \label{eq:23}
 \end{equation}
we have
\begin{eqnarray}
A(G_{\parallel}=0)=\frac{2\sqrt{\pi}\,\eta}{S}\sum_{ij}d_{z}
(\mathbf{r}_{i})d_{z}(\mathbf{r}_{j})\nonumber\\
=\frac{2\sqrt{\pi}\,\eta}{S}\sum_{n}(D_{z}^{n})^{2},
 \label{eq:24}
\end{eqnarray}
where $D_{z}^{n}$ is the $z$-component of the total dipole moment
in the $n$-th layer parallel to the $(x,y)$ plane.

Finally we note that in both 1D and 2D cases the dipole energy
(\ref{eq:4}) can be rewritten  as
\begin{equation}
\mathcal{E}_{dip}= \frac{1}{2}\sum_{ \alpha\beta, ij }
Q_{\alpha\beta,ij}d_{\alpha}(\mathbf{r}_{i})d_{\beta}(\mathbf{r}_{j}),
\label{eq:25}
\end{equation}
where $Q_{\alpha\beta,ij}$ is the structure constant matrix; the
latter  can be calculated once and for all similar to the case of
3D periodicity\cite{zhong}. Explicitly, the matrices are
\begin{widetext}
\begin{eqnarray}
&&Q_{\alpha\beta,ij}^{(1D)}=\frac{2}{a}\sum_{\mathbf{G}}
G^{2}\,\cos(\mathbf{G}\cdot \mathbf{z}_{ij})\nonumber
\bigg\{K_{0}\big(G\rho_{ij}\big) \: \delta_{\alpha z}\delta_{\beta
z} \nonumber +\frac{\delta_{\alpha x}\delta_{\beta x}
+\delta_{\alpha y}\delta_{\beta y}}{G\,\rho_{ij}}
K_{1}\big(G\rho_{ij}\big)\nonumber\\
&&-\frac{1}{\rho_{ij}^2}K_{2}\big(G \rho_{ij}\big)\,
\rho_{\alpha,ij}\,\rho_{\beta,ij}\bigg\}\nonumber
-\frac{2}{a}\sum_{\mathbf{G}}\: G\,\sin(\mathbf{G}\cdot
\mathbf{z}_{ij})
\:K_{1}\big(G\rho_{ij}\big)\,\rho_{ij}^{-1}\nonumber \times
G_{\alpha} \rho_{\beta,ij}+ \nonumber\\
&&+\frac{1}{a^{3}}\big(\delta_{\alpha x}\delta_{\beta x}+
\delta_{\alpha y}\delta_{\beta y}- 2\delta_{\alpha z}\delta_{\beta
z}\big)\sum_{n=-\infty}^{\infty\quad\prime}
\bigg|n+\frac{z_{ij}}{a}\bigg|^{-3},
\nonumber\\
&&Q_{\alpha\beta,ij}^{(2D)}=\frac{2\pi}{S}\sum_{\mathbf{G}}
\bigg\{G\cos(\mathbf{G}\cdot \boldsymbol{\rho}_{ij})\bigg
[\frac{1}{\sqrt{4\pi}}\:\Gamma\left(
 -\frac{1}{2},\,\frac{G^{2}}{4\eta^{2}}
 \right)\delta_{\alpha z}\delta_{\beta z}
 +\frac{1}{G^2}\:\text{erfc}\left(
 \frac{G}{2\eta}
 \right)G_{\alpha}G_{\beta}\bigg ]
\nonumber\\
 &&+G\,
\exp(-G\,|z_{ij}|)\bigg[\big(\frac{G_{\alpha}G_{\beta}}{G^2}
-\delta_{\alpha z}\delta_{\beta
z}\big)\cos(\mathbf{G}\cdot\boldsymbol{\rho}_{ij})
\,-\frac{G_{\alpha}\delta_{\beta
z}}{G}\sin(\mathbf{G}\cdot\boldsymbol{\rho}_{ij})
\frac{z_{ij}}{|z_{ij}|}\bigg]\bigg\}
-\frac{4\eta^{3}\,\delta_{\alpha \beta}\delta_{ij}}{3\sqrt{\pi}}.
\nonumber\\
 \label{eq:26}
\end{eqnarray}
\end{widetext}

\end{document}